\documentclass[12pt]{article} 

\begin{document} 
  \vspace*{1.1cm} 
  \begin{center} 
  {\LARGE \bf Inertial Mass and Vacuum Fluctuations \\
  in Quantum Field Theory} 
  \end{center} 

  \begin{center} 
  \vskip 10pt 
  Giovanni Modanese
  \vskip 5pt
  {\it California Institute for Physics and Astrophysics \\
  366 Cambridge Ave., Palo Alto, CA 94306}
  \vskip 5pt
  and
  \vskip 5pt
  {\it University of Bolzano -- Industrial Engineering \\
  Via Sernesi 1, 39100 Bolzano, Italy}
  \vskip 5pt
  E-mail address: giovanni.modanese@unibz.it  
  \end{center} 

  \baselineskip=.175in 
    
\begin{abstract}
Motivated by recent works on the origin of inertial mass, we revisit the relationship between the mass of charged particles and zero-point electromagnetic fields. To this end we first introduce a simple model comprising a scalar field coupled to stochastic or thermal electromagnetic fields. Then we check if it is possible to start from a zero bare mass in the renormalization process and express the finite physical mass in terms of a cut-off. In scalar QED this is indeed possible, except for the problem that all conceivable cut-offs correspond to very large masses. For spin-1/2 particles (QED with fermions) the relation between bare mass and renormalized mass is compatible with the observed electron mass and with a finite cut-off, but only if the bare mass is not zero; for any value of the cut-off the radiative correction is very small.

\medskip



\end{abstract}

\bigskip \bigskip

In modern physics each elementary particle is characterised by a few parameters which define essentially its symmetry properties. Mass and spin define the behavior of the particle wavefunction with respect to spacetime (Poincar\'e) transformations; electric charge, barion or lepton number etc. define its behavior with respect to gauge transformations. These same parameters also determine the (gravitational or gauge) interactions of the particle.

Unlike spin and charge, mass is a continuous parameter which spans several magnitude orders in a table of the known elementary particles. In spite of several attempts, there is no generally accepted way of expressing these masses, or at least their scale, in terms of fundamental constants. In the standard model particles acquire a mass thanks to the Higgs field, but the reproduction of the observed spectrum is only possible by choosing a different coupling for each particle.

Inertia in itself is not really explained by quantum field theory; rather, it is incorporated in its formalism as an automatic consequence of the spacetime invariance of the classical Lagrangians. In turn, these Lagrangians are essentially a generalization of Newtonian dynamics. In the equations for quantum fields, like in the wave equations for single particles or in their classical limits, mass appears as a free parameter which can take zero or positive values.

Therefore it is not surprising that several works in the last years (for a discussion and a list of references see for instance \cite{Rosu}) have been devoted to the search of a possible fundamental explanation of the inertial properties of matter. Some of these works look for the source of inertia in the interaction between charged particles and the electromagnetic vacuum fluctuations, exploring analogies with the dynamical Casimir forces on an accelerated cavity \cite{Jaekel} or with the unbalanced radiation pressure in the Davies-Unruh thermal bath seen in accelerated frames \cite{Haisch1}. The possibility was also investigated, in connection with astrophysical problems, that Newton law does not hold true for very small acceleration \cite{Milgrom}.

In this work, we try to clarify whether some of the proposals contained in the mentioned papers can be implemented, or at least partially analysed, within the standard formalism of quantum field theory--perhaps leading to a more satisfactory inclusion of the concept of mass. Of course, the idea of dynamical mass generation induced by vacuum fluctuations is already familiar in quantum field theory \cite{Coleman}, but it is usually connected to phenomena of spontaneous symmetry breaking, where a quantum field acquires a non-zero vacuum expectation value. Here, on the other hand, we are interested only into the effects of the fluctuations.

One should also keep in mind that the mass of a particle can come into play, in quantum field theory, in different equivalent forms, namely as (a) the response to the coupling with an external field; (b) a parameter in the dispersion relation $E(k)$; (c) the pole in the particle propagator and in its creation/annihilation cross section.

The pragmatic attitude of quantum field theory towards the origin of mass curiously seems to disappear only at one point, namely when in the mass renormalization procedure the ``bare" mass $m_0$ is assumed to be infinite. What happens  if we introduce finite cut-offs in the field theoretical expressions for the radiatively induced mass shift $\Sigma$, and set $m_0=0$? One finds that the result depends much on the spin of the particles. For scalar particles, it is possible to introduce a cut-off in $\Sigma$, set the bare mass to zero and interpretate somehow the physical mass as entirely due to vacuum fluctuations--except for the problem that the ``natural" cut-offs admitted in quantum field theory (supersymmetry scale, GUT scale, Planck scale) all correspond to very large masses. For spin-1/2 particles (QED with fermions) one obtains a relation between bare mass and renormalized mass which is compatible with the observed electron mass and with a finite cut-off, but only if the bare mass is not zero.  Below we shall give the explicit expressions for the scalar and spinor case. Before that, however, it is useful to consider a semiclassical approximation, which turns out to contain much of the physics of the problem.

Let us consider charged particles with bare mass $m_0$ immersed in a thermal or 
stochastic background $A_\mu(x)$. For scalar particles described by a quantum field $\phi$, the Lagrangian density is of the form 
 	\begin{equation} 
	L = \frac{1}{2} \phi^* (P^\mu - e A^\mu)
	\phi (P_\mu - e A_\mu)
	- \frac{1}{2} m_0^2 |\phi|^2
\label{lagr}
\end{equation}
and contains a term $e^2 \phi^* A_\mu A^\mu \phi$, which after averaging on $A_\mu$ can be regarded as a mass term for the field $\phi$. Take, for instance, the 
Coulomb gauge: The effective squared mass turns out to be equal to $m^2 = m_0^2 + e^2 \langle | {\bf A}({\bf x}) |^2 \rangle$. 
  
For homogeneous black body radiation at a given temperature $T$, the average is readily computed. One has 
\begin{equation} 
\langle | {\bf A} |^2 \rangle = 
\int_0^\infty d\omega \frac{u_\omega}{\omega^2} 
\label{ave} 
\end{equation} 
where $u_\omega$ is the Planck spectral energy density. By integrating one finds that the squared mass shift is given by $\Delta m^2 = const. \sqrt{\alpha} k_BT$ (the 
constant is adimensional and of order 1). This mass shift can be significant in a hot plasma, but only for spin-zero particles, not for fermions. In fact, the Dirac Lagrangian is linear with respect to the field $A_\mu$, therefore it is impossible to obtain a mass term for spinors by averaging over the electromagnetic field. One expects that a mass shift for fermions will only appear at one-loop order. This is in fact confirmed by the full calculation in thermal quantum field theory \cite{LeBellac} and by experimental evidence (no relevant mass shifts are observed in the Sun). Note that although a second-order formalism for Dirac fermions in QED exists, it has been used until now for calculations with internal fermion lines only \cite{Morgan}. The result above seems to confirm that a proper treatment of on-shell fermions intrinsically requires a first-order lagrangian.

Eq.\ (\ref{ave}) can also be applied to the Lorentz-invariant frequency spectrum of the zero-point field in Stochastic Electrodynamics, namely $u_\omega=\omega^3$ \cite{Milonni}. In this case the integral diverges, unless we introduce either a cut-off, or a resonant coupling of the zero-point field to the particle at a certain frequency $\omega_0$--which therefore defines the mass of the particle \cite{Haisch2}. This could be viewed as an alternative to mass generation by coupling to the Higgs, but, again, only for scalar particles.

Turning now to scalar QED, one can consider the Feynman mass renormalization condition $m_0^2+\Sigma(m^2)=m^2$, set $m_0=0$, impose a physical cut-off $M$ in $\Sigma$ and compute $m$ as a function of $M$. One finds in this way, as mentioned, that $m$ is of the order of the cut-off. Actually, scalar QED only describes particles like pions or other charged mesons which are not regarded as fundamental. Better known and physically more relevant is spinor QED, i.e. the quantum electrodynamics of spin $1/2$ charged particles. Perturbative expansions in spinor QED lead to some divergences in the radiative corrections, but such divergences are usually mild ones. Starting in the Sixties, the limit in which spinors have zero bare mass has been studied in great detail, and some general theorems were proven. Novel infrared divergences appear in this limit, but as first shown in \cite{nl}, under certain physical conditions all transition matrix elements are finite. The zero mass limit is furthermore important because it corresponds to the limit in which the charged particles are not massless, but interaction energies are very large compared to the mass scale \cite{jb}. This ultraviolet limit is governed by Weinberg theorem \cite{iz}, which predicts the behaviour of transition amplitudes in the limit of large external momenta.
 
The authors of \cite{nl} wondered if charged particles can have zero mass. From the classical point of view this looks impossible, because the energy of the electric field generated by a particle bears a mass--the well known electromagnetic mass of the electron. In the quantum theory, however, this is not obvious a priori. An analysis of the renormalization of radiative corrections to the self mass is required. The first step in this direction was taken by the authors of \cite{Baker1}, who investigated whether, more generally, a finite electron mass could be generated by radiative corrections starting from zero bare mass and found that a full radiatively-generated mass was possible, provided the photon wave function renormalization constant $Z_3$ was finite. Later, they established the fact that $Z_3$ has, perturbatively, logarithmic divergences to all orders \cite{Baker2}. They concluded that the only way to arrive at a finite $Z_3$ is to make the prefactor of this logarithm vanish, i.e. in modern language, to find a nontrivial zero of the QED beta function. Up to now, however, there are no hints of the existence of such a zero.

In the modern picture of elementary particles the historical results of spinor QED are generalized to include those of unified electroweak theory and, in principle, the strong interactions for hadrons (which, however, do not admit a perturbative treatment except in special cases). Disregarding quark-gluons interactions and without going into details which fall outside the scope of this work, we just observe here that the masses of all fermions in the electroweak theory are subjected to radiative corrections due to the electromagnetic field and to the field of the W and Z vector bosons. Being vector bosons massive, their radiative corrections are far less divergent; in particular, they are compatible with zero mass neutrinos.

All the above refers to quantum field theory renormalized in the usual sense, i.e. in the limit where all energy cut-offs tend to infinity. This approach has been highly successful in explaining the phenomenology of elementary particles. Renormalizability of the theory implies that phenomena occurring at very high energies do not affect results at the energy scale of interest. In matters of principle, however--like the origin of inertia considered in this paper--one can take a different attitude. One can admit that an intrinsic high-energy cut-off for quantum field theory exists, due to some more general high energy theory or to quantum gravity (the Planck scale). It is therefore interesting to see what QED predicts in this case. Let us consider the simplest radiative correction to the self mass, namely the one-loop self energy diagram, and the corresponding expression for the mass renormalization condition, that is
	\begin{equation} 
	m-m_0 = \left[ \Sigma(p^2,M) 
	\right]_{\sqrt{p^2}=m} = m_0 \frac{3\alpha}{4\pi}
	\left( \ln \frac{M^2}{m_0^2} + \frac{1}{2}
	\right)
\label{mild}
\end{equation}

Inserting $m_0=0$ we find that the radiatively corrected mass $m$ is zero, too. A charged massless particle could then exist, under the assumption of a finite cut-off, thanks to the mild logarithmic divergence. If we set $m_0 > 0$ instead, we obtain a radiative correction which even for very large values of the cut-off is much smaller than the electron scale. In fact, let us parametrize the cut-off as $M=10^\xi$, set $m/m_0 \equiv k >1$ (i.e., vacuum fluctuations increase the mass by a factor $k$), solve eq.\ (\ref{mild}) for $\xi$ and plot the inverse function $k(\xi)$ (Fig.\ 1). We see that even for very large values of the cut-off, the renormalization effect is quite moderate. 

In conclusion, in the framework of quantum field theory it is impossible to interpret the observed mass of the electron as due to radiative corrections. Namely

(1) in the renormalized theory, the radiative corrections to mass are divergent if the bare mass is zero;

(2) in the presence of a large but finite energy cut-off, the one-loop radiative correction vanishes for zero mass and is otherwise irrelevant.

\bigskip
\noindent
{\bf Acknowledgments} - This work was supported in part by the California Institute for Physics and Astrophysics via grant CIPA-MG7099. I am grateful to V.\ Hushwater, V. Savchenko and A.\ Rueda for useful discussions. I also thank C.\ Schubert for bringing to my attention ref.s \cite{Baker1}, \cite{Baker2} and their meaning in terms of the renormalization group equation.

\newpage

\noindent {\bf Figure caption}

\noindent Fig.\ 1 - Ratio between renormalized mass and bare mass in QED with an UV cut-off. The graph is obtained by solving eq.\ (\ref{mild}) numerically.

\end{document}